\documentclass[aps,prd,showpacs,preprint,amsmath,amssymb]{revtex4}

\usepackage{bm}
\usepackage {graphicx}

\begin{document}


\title{Perturbative Study of Bremsstrahlung and Pair-Production by \\ Spin-$\frac{1}{2}$ Particles in the Aharonov-Bohm Potential}


\author{U. A. al-Binni and M. S. Shikakhwa}
\affiliation{Department of Physics, University of Jordan, Amman
111942, Jordan}



\begin{abstract}

In the presence of an external Aharonov-Bohm potential, we
investigate the two QED processes of the emission of a
bremsstrahlung photon by an electron, and the production of an
electron-positron pair by a single photon. Calculations are
carried out using the Born approximation within the framework of
covariant perturbation theory to lowest non-vanishing order in
$\alpha$. The matrix element for each process is derived, and the
corresponding differential cross-section is calculated. In the
non-relativistic limit, the resulting angular and spectral
distributions and some polarization properties are considered, and
compared to results of previous works.

\end{abstract}


\maketitle

\section{Introduction}
The Aharonov-Bohm (AB) effect \cite{ahar59} is one of the
remarkable predictions of quantum theory. It has shown that,
contrary to classical electrodynamics, enclosed electromagnetic
fields can interact with charged quantum particles via the vector
potential leading to observable effects, even in the absence of
Lorentz forces. Comprehensive reviews covering many aspects of the
subject are \cite{olar85,peshk89}.

Traditionally, the AB effect has been studied in connection with
elastic scattering and bound state problems. In recent years,
however, a limited number of works have been devoted to processes
that go beyond these problems and that open the way for a deeper,
more detailed understanding of the AB effect. The two processes
that have been addressed by these works were the emission of a
single bremsstrahlung photon by a charged particle, and the
production of particle-antiparticle pair in the AB potential. The
first of these works \cite{sereb88}, investigated the emission of
bremsstrahlung by a spinless non-relativistic particle by looking
at the spectral and angular distribution of emitted photons. Then
the same process was studied relativistically in \cite{gal90}
using Klein-Gordon particles, this time the polarization of the
emitted photon was taken into account, and both non-relativistic
and ultra-relativistic limits of the results were considered. For
relativistic spin-$\frac{1}{2}$ electrons, the process was
investigated in reference \cite{aud96} for Dirac particles, where
the differential cross-section and its limiting cases were used to
discuss spectral and angular distributions of emitted photons.
That study included electron spin polarization and its effect on
the polarization of the emitted photon, and angular momentum
selection rules were reported.

As for the process of pair-production in the AB potential, only
two studies have been conducted: the first was \cite{skar96},
where the differential cross-section was found and the effect of
photon polarization on spin polarization of created particles was
considered. Here also, angular momentum selection rules for
angular momentum were reported. The case of spinless
pair-production has been recently studied in \cite{sha02}, where
the photon-polarized differential cross-section was calculated,
and selection rules found.

In all of these works, the method of finding the scattering
amplitude uses \emph{exact wavefunctions} of the Klein-Gordon or
Dirac particles in the AB potential, and then the scattering
matrix is calculated to first order using scattering states
constructed from these functions.

The aim of the present paper is to study the processes of
bremsstrahlung and pair-production in an external AB potential in
the Born approximation using covariant perturbation theory to the
lowest non-vanishing order, which corresponds in this case to
$\mathcal{O}(\alpha^2)$. The present work was motivated by the
series of works by Skarzhinsky \emph{et al.}
\cite{aud96,skar96,aud98}, and in that respect, the motivation was
twofold: first, we think it is interesting, as such, to treat the
problem using this method, as it is the one traditionally employed
in this kind of calculation, where results are cast into the
familiar Bethe-Heitler form, and to our knowledge, no such
calculation has been reported for the AB potential. Another
reason, is to test the applicability of the Born approximation to
this particular problem by comparing with the results of the
``exact'' approach, especially since there were some speculations
regarding the applicability of the small field approximation to
this problem \cite{skar97,aud98}.

We will use the idealization usually followed in describing the
AB-potential, namely to assume it is generated by a very thin,
very long flux tube putting out a potential given by:
\begin{equation}
\label{2}A^{\mu}(x)=\frac{\Phi}{2\pi (x^2+y^2)}(0,-y,x,0)
\end{equation}
so that the direction of magnetic field generating the flux $\Phi$
is along the $z$-axis. In terms of the magnetic flux quantum
$\Phi_0=2\pi / \left| e \right|$, we can write
$\Phi/\Phi_0\equiv\delta + N$, where $N$ is the integral part and
$\delta$ is the remaining fraction. To use the Born approximation,
we will assume the flux to be small enough, so that $N=0$ and
$\delta\ll 1$.

This paper is organized into four sections: following this
introductory section, we present in section II the calculation of
the matrix element and differential cross-section for the emission
of a bremsstrahlung photon from a Dirac particle in the
AB-potential. We also calculate the differential cross-section in
the non-relativistic limit.

In section III we conduct an analogous study for the process of
production of a Dirac pair in the AB-potential, where the matrix
element is found and the corresponding differential cross-section
is calculated. The limit of low energy photon is found as well,
and the polarization effects are investigated.

Section IV summarizes our results and states the conclusions and
the problems that are still open.

As regards conventions, we follow those of ref. \cite{Itz80},
specifically, we use the Pauli-Dirac representation of the
$\gamma$-matrices.

\section{Bremsstrahlung in the AB potential}
\subsection{Amplitude and Cross-section}
The probability amplitude for the emission of a single
bremsstrahlung photon is given by the matrix element:
\begin{equation}
\label{3} S_{fi}=\langle e^{-}\gamma|\hat{S}|e^{-}\rangle
\end{equation}
The two diagrams that contribute to the process at tree-level are
shown in figure (\ref{fig1}), their sum gives us the amplitude to
lowest non-vanishing order in $\alpha$ ($\mathcal{O}(\alpha^2)$):
\begin{figure}[htbp]
\includegraphics[width=8cm]{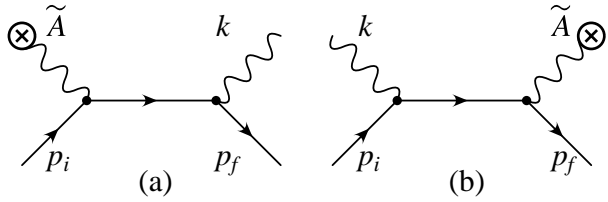}
\caption{\label{fig1} Tree-level diagrams for bremsstrahlung.}
\end{figure}
\begin{eqnarray}
\label{4} S^{(2)}_{fi}=-ie^{2}\bar{u}^{(\beta)}(p_f)\Big[\not\!
\epsilon^{(\lambda)*}\frac{\not\! p_f + \not\! k +m}{2p_f\cdot k}
\not\!\! \tilde{A}^{\mathrm{ext}}(q)-
\not\!\! \tilde{A}^{\mathrm{ext}}(q)\frac{\not\! p_i - \not\! k
+m}{2p_i\cdot k}\not\! \epsilon^{(\lambda)*}\Big]u^{(\alpha)}(p_i)
\end{eqnarray}
where $u^{(s)}(p)$ is a 4-component spinor satisfying the free
Dirac equation for a particle with 4-momentum $p$ and in a spin
state $s$. $p_i$, $p_f$, and $k$ refer to the 4-momenta of the
incoming and outgoing Dirac particles, and the photon,
respectively. $\epsilon_{\mu}^{(\lambda)}$ is the photon
polarization vector, and $\tilde{A}_{\mu}^{\mathrm{ext}}(q)$ is
the Fourier transform of the AB-potential:
\begin{equation}
\label{5}\tilde{A}_{\mu}^{\mathrm{ext}}(q)=-\frac{(2\pi)^2i\Phi}{\left|\mathbf{q}\right|^2}\delta
(q^0)\delta (q^3)R_\mu
\end{equation}
Here, the momentum transfer 4-vector $q^\mu$ is given by:
\begin{equation}
\label{6}q^{\mu}=p_{f}^{\mu}+k^{\mu}-p_{i}^{\mu}
\end{equation}
and $R^\mu$ is a vector defined by:
\begin{equation}
\label{7}R^{\mu}\equiv\big(0,\hat{z}\times\mathbf{q}\big)=(0,-q^2,q^1,0)
\end{equation}
The nature of the problem implies that the energy is conserved
($q^0=0$) and that the momentum along the flux tube's axis is also
conserved ($q^3=0$), whereas the radial component $q^\bot$ is not,
and momentum must be transferred to the tube. These considerations
are expressed in the two $\delta$-functions appearing in
(\ref{5}). Now, writing (\ref{4}) with the help of (\ref{5}), we
get:
\begin{eqnarray}
\label{8}
\begin{aligned}
S_{fi}^{(2)}& \equiv\Lambda\bar{u}^{(\beta)}(p_f)F^{(\lambda)}u^{(\alpha)}(p_i)\\ 
& =\Lambda\bar{u}^{(\beta)}(p_f)\Big[\not\!
\epsilon^{(\lambda)*}\frac{\not\! p_f + \not\! k +m}{2p_f\cdot k}\not\!\! R
-\not\!\! R\frac{\not\! p_i - \not\! k +m}{2p_i\cdot k}\not\!
\epsilon^{(\lambda)*}\Big]u^{(\alpha)}(p_i)
\end{aligned}
\end{eqnarray}
where $\Lambda\equiv -(2\pi e)^2\Phi\delta (q^0)\delta
(q^3)/\left|{\bf{q}}\right|^2$, and the indices $\alpha$, $\beta$
and $\lambda$ stand for polarization states of the incoming and
outgoing Dirac particles and the photon, respectively. The
differential cross-section per unit length of the solenoid
resulting from this amplitude, as calculated according to the
conventions of \cite{Itz80}, reads:
\begin{eqnarray}
\label{9}d\sigma_{(\alpha\rightarrow\beta;\lambda)}=\frac{|\Lambda|^2}{|\mathbf{p}_{i}|/m}
\Big(\frac{md^3p_f}{p_{f}^{0}(2\pi)^3}\Big)\Big(\frac{d^3k}{2\omega
(2\pi)^3}\Big)
\operatorname{tr} \left\{ {\mathcal{P} _i^{(\alpha)} \bar
F^{(\lambda)}\mathcal{P} _f^{(\beta)} F^{(\lambda)}} \right\}
\end{eqnarray}
Here $\mathcal{P}^{(s)}=u^{(s)}(p)\bar{u}^{(s)}(p)$ is an
appropriate projection operator, $F^{(\lambda)}$ is as defined in
(\ref{8}), and
$\bar{F}^{(\lambda)}=\gamma^{0}F^{(\lambda)\dag}\gamma^{0}$. We
will be using the formula in (\ref{9}) twice; once to evaluate the
differential cross-section for unpolarized electrons, and once for
polarized electrons. The calculations are straightforward, but
quite lengthy, especially in the polarized case. The traces in
either case were evaluated with the help of a computer programme
called FeynPar \cite{west93}, which runs on Mathematica. The raw
output of FeynPar was then rearranged by hand to become compact
and physically transparent.

In the unpolarized case $\mathcal{P}^{(s)}$ projects over positive
energy states, whereas in the polarized case it further projects
over states with well-defined spin polarization. For the former
the effect of polarization is cancelled out by summing over states
of the outgoing particle and averaging over those of the incoming
particle, and the resulting projector is well-known (e.g. see
\cite{Itz80}):
\begin{equation}
\label{9.5} \bar{\mathcal{P}}= \sum\limits_s {u^{\left( s \right)}
\left( p \right)\bar u^{\left( s \right)} \left( p
\right)}=\frac{(\not\! p+m)}{2m}
\end{equation}

Plugging this projector into (\ref{9}), and integrating out the
$\delta$-functions, we get an unpolarized cross-section that is
differential in the azimuthal angle of the outgoing electron
$\varphi_f$, the photon momentum
$\left|\mathbf{k}\right|\equiv\omega$, and the solid angle for the
photon $\Omega_\gamma$:
\begin{subequations}
\label{11}
\begin{equation} {\frac{{d\bar{\sigma} _{(\lambda )} }} {{d\varphi _f
d\omega d\Omega _\gamma }}}=A+B  \label{subeq:11a}
\end{equation}
where:
\begin{equation}
\begin{aligned}
 A& = \frac{{2e^2 (\pi \delta )^2 \omega }}
{{(2\pi )^4 \left| {{\mathbf{p}}_i } \right|\left| {\mathbf{q}}
\right|^4 }}\left| {\frac{{(\epsilon ^{(\lambda )} \cdot p_f )(R
\cdot p_i )}}
{{p_f \cdot k}}} \right. 
 \left. { - \frac{{(\epsilon ^{(\lambda )} \cdot p_i )(R \cdot p_f )}}
{{p_i \cdot k}} - R \cdot \epsilon ^{(\lambda )} } \right|^2 \hfill \\
  B& = \frac{{e^2 (\pi \delta )^2 \omega }}
{{2(2\pi )^4 \left| {{\mathbf{p}}_i } \right|}}\left[ {\left|
{\frac{{\epsilon ^{(\lambda )}  \cdot p_f }} {{p_f  \cdot k}} -
\frac{{\epsilon ^{(\lambda )}  \cdot p_i }}
{{p_i  \cdot k}}} \right|^2 } \right. 
  \left. { + \frac{{\left( {k^ \bot  } \right)^2 }}
{{(p_f  \cdot k)(p_i  \cdot k)}}} \right] 
\end{aligned}
\label{subeq:11b}
\end{equation}
\end{subequations}
$k^{\bot}$ being the radial part of $\mathbf{k}$, and the bar over
$\sigma$ means that this stands for the unpolarized cross-section.

To find the polarized differential cross-section, we need to use
concrete wavefunctions that describe particles in well-defined
states of spin polarization. To this end, we chose wavefunctions
that are eigencvectors of the 3rd component of the spin
pesudotensor $\hat{S}^\mu$ \cite{sok86}:
\begin{equation}
\label{12} {\hat S^3 = \beta \Sigma ^3 + \frac{{p^3 }} {m}\gamma
^5 }
\end{equation}
where, $\beta$, $\Sigma^{3}$ and $\gamma^{5}$ are the usual
matrices, to be used in our case in the Pauli-Dirac
representation, and $p^3$ is the $z$-component of the momentum
vector. This operator has the benefit of reducing to the usual
spin projection in the non-relativistic limit. It has also been
used in the works with which we shall be comparing our results
\cite{aud96,skar96}. The free Dirac wavefunction $psi$ that can be
used to construct the projector $\mathcal{P}$ will satisfy the
following combined eigenvalue problem:
\begin{equation}
\label{13}
\begin{array}{*{20}c}
 {\hat H\psi = E\psi ,} & {\hat p^3 \psi = p^3 \psi ,} & {\hat S^3 \psi = s\psi .} \\
 \end{array}
\end{equation}
where $\hat H$ is the free Dirac Hamiltonian, and $s$, the
eigenvalue of $\hat{S}^3$, which takes the values $\pm \sqrt {1 +
\left( {p^3 /m} \right)^2 }$. The desired functions (with
normalization $ \int {\bar \psi _j \psi _{j'} } = \delta _{jj'}$)
are:

\begin{subequations}
\label{14}
\begin{equation} \psi _j (p) = e^{ - ip \cdot x} u_j \label{subeq:14a}
\end{equation}
where the spinor part $u_j$ is given by:
\begin{equation}
u _j (p) = \frac{1} {{2\sqrt {ms} }}\left( {\begin{array}{*{20}c}
 {\sqrt {E + sm} \sqrt {s + 1} } \\
 {\epsilon _3 \sqrt {E - sm} \sqrt {s - 1} e^{i\varphi _p } } \\
 {\epsilon _3 \sqrt {E + sm} \sqrt {s - 1} } \\
 {\sqrt {E - sm} \sqrt {s + 1} e^{i\varphi _p } } \\
 \end{array} } \right)
\label{subeq:14b}
\end{equation}
\end{subequations}
in this equation, $\epsilon
_3=\operatorname{sgn}(s)\operatorname{sgn}(p^3)$, and $j$ stands
for the set of quantum numbers $(E,p^3,s)$, and the resulting
projection operator then becomes:
\begin{equation}
\label{15} u_j\bar{u}_j\equiv{\mathcal{P}}_{+}^{(s)}(p)= \frac{1}
{{4sm}}\left( {\not\! p + m} \right)\left( {s - \gamma ^5 \gamma
^3 - \frac{{p^3 }} {m}\gamma ^5 } \right)
\end{equation}
The polarized cross-section is obtained by plugging (\ref{15})
into (\ref{9}). This calculation is greatly facilitated if we
rewrite $F^{(\lambda)}$ appearing in (\ref{8}) in terms of the
following quantities (in which the index $(\lambda)$ is omitted
for brevity):
\begin{subequations}
\label{16}
\begin{equation}
\begin{array}{*{20}c}
 {V^\mu \equiv a\epsilon ^{\mu *} + \left({b-c}\right)\eta ^{\mu *},}&{U\equiv -\left({b+c}\right)\epsilon ^{3*}},\\
 \end{array}\label{subeq:16a}
\end{equation}
where:
\begin{align}
a& \equiv\frac{R\cdot p_i}{p_{f}\cdot k}-\frac{R\cdot
p_f}{p_{i}\cdot k},& b& \equiv
-\frac{|\mathbf{q}|^2}{2}\frac{1}{p_f\cdot k},\nonumber\\
c& \equiv \frac{|\mathbf{q}|^2}{2}\frac{1}{p_i\cdot k},&
{\eta}^{\mu}& \equiv(0,\hat{z}\times\boldsymbol{\epsilon})=(
0,-{\epsilon}^2,{\epsilon}^1,0). \label{subeq:16b}
\end{align}
\end{subequations}
Doing this, we end up with the following:
\begin{subequations}
\label{17}
\begin{equation}
\operatorname{tr} \left\{ {\mathcal{P} _i^{(s)} \bar
F^{(\lambda)}\mathcal{P} _f^{(r)} F^{(\lambda)}} \right\} =
\frac{{1 }} {{16rsm^2 }}\left( {J_1  + J_2  + J_2^*  + J_3 }
\right) \label{subeq:17a}
\end{equation}
where
\begin{equation}
\begin{split}
 J_1& = 4\left\{ {\left[ {\left( {V^* \cdot p_f } \right)\left( {V \cdot p_i } \right) + \left( {V^* \cdot p_i } \right)\left( {V \cdot p_f } \right)} \right.} \right.\\
  & \quad \left. { + V^* \cdot V\left( {m^2 - p_i \cdot p_f } \right)} \right]\left( {1 + rs + \frac{{p_f^3 p_i^3 }}
{{m^2 }}} \right) + V^* \cdot V\left( {k^3 } \right)^2 \\
& \quad +\frac{i}{m}\epsilon _{\alpha \beta \gamma \delta }
p_f^\alpha p_i^\beta V^{\gamma *} V^\delta \left( {sp_f^3 + rp_i^3
} \right) - im\omega \left( {r + s} \right)\left( {{\mathbf{V}}^* \times {\mathbf{V}}} \right) \\
& \quad + 2\left| {V^3 } \right|^2 \left( {m^2 - p_i \cdot p_f }
\right) \left. { - k^3 \left[ {V^{3*} V \cdot \left( {p_i - p_f } \right) + V^3 V^* \cdot \left( {p_i - p_f } \right)} \right]} \right\} \\
\end{split}
\label{subeq:17b}
\end{equation}
\begin{equation}
\begin{split}
  J_2&  = 4U\left\{ {\frac{i}
{m}\left( {rp_i^3  + sp_f^3 } \right)\left( {p_i^0 V^*  \cdot p_f
+ p_f^0 V^*  \cdot p_i } \right)} \right. - rs\left( {{\mathbf{p}}_i  \times {\mathbf{p}}_f } \right) \cdot {\mathbf{V}}^*  \\
& \quad + p_f^3 s^2 \left( {{\mathbf{V}}^*  \times {\mathbf{p}}_f } \right) \cdot \hat z - p_i^3 r^2 \left( {{\mathbf{V}}^*  \times {\mathbf{p}}_i } \right) \cdot \hat z \\
& \quad  \left. { + V^{3*} \left[ {im\omega \left( {r - s} \right)
+ \left( {1 - \frac{{p_f^3 p_i^3 }}
{{m^2 }}} \right)\left( {{\mathbf{p}}_i  \times {\mathbf{p}}_f } \right) \cdot \hat z} \right]} \right\} \hfill \\
\end{split}
\label{subeq:17c}
\end{equation}
\begin{equation}
\begin{split}
  J_3& = 4\left| U \right|^2 \left\{ {\left( {rs - 1 + \frac{{p_i^3 p_f^3 }}
{{m^2 }}} \right)\left[ {p_i  \cdot p_f  + 2{\mathbf{p}}_i  \cdot
{\mathbf{p}}_f  - m^2 } \right]} \right. \left. { + \left( {p_f^3
+ p_i^3 } \right)^2 } \right\}
\end{split}
\label{subeq:17d}
\end{equation}
\end{subequations}
In these expressions $s$ and $r$ stand for the pseudo-spin
eigenvalues for the incoming and outgoing particles, respectively.
We can simplify (\ref{17}), without loss of generality, by
assuming the electron to be incident normally on the solenoid,
i.e.
\begin{equation}
\label{18} p_{i}^{3}\equiv 0~\Rightarrow~p_{f}^{3}= -k^{3}
\end{equation}
As a result of this condition, $s$ takes on the values $\pm 1$,
thereby exhibiting a behaviour much like the usual spin
projection. Further simplification is possible by an appropriate
choice of the polarization vectors for the emitted photon. We will
use the following vectors, which are constructed in a way that
takes advantage of the direction of the magnetic field in the tube
\cite{sok86}:
\begin{equation}
\label{19}
\begin{split}
\epsilon ^{\left( \sigma  \right)}&  = \left( {0, - \sin \varphi
_k ,\cos \varphi _k ,0} \right) \\
\epsilon ^{\left( \pi  \right)}&  = \left( {0, - \cos \theta _k
\cos \varphi _k , - \cos \theta _k \sin \varphi _k ,\sin \theta _k
} \right)
\end{split}
\end{equation}
where $\theta_k$ and $\varphi_k$ are the polar and azimuthal
angles of the the photon's momentum vector, respectively. From
these linear polarization vectors we can construct vectors
describing circular polarization: $\epsilon ^{\left( \ell \right)}
= \left( {\epsilon ^{\left( \sigma \right)} + i\ell \epsilon
^{\left( \pi \right)} } \right)/\sqrt 2$.

In terms of the unpolarized cross-section as given in (\ref{11}),
the polarized cross-section reads now:
\begin{equation}
\label{20}
\begin{split}
  \frac{{d\sigma _{\left( {s  \to r ,\lambda } \right)} }}
{{d\omega d\Omega _\gamma  d\varphi _f }}& = \frac{{1 + rs}}
{{2rs}}\left( {\frac{{d\bar \sigma _{\left( \lambda  \right)} }}
{{d\omega d\Omega _\gamma  d\varphi _f }}} \right) + \frac{{e^2
(\pi \delta )^2 \omega }} {{(2\pi )^4 sr\left| {{\mathbf{p}}_i }
\right|\left| {\mathbf{q}} \right|^4 }}\left\{ {\tfrac{1}
{2}\left( {V \cdot V} \right)\left( {k^3 } \right)^2 } \right. \\
& \quad  + \left( {V^3 } \right)^2 \left( {m^2  - p_i  \cdot p_f } \right) - k^3 V^3 V \cdot \left( {p_i  - p_f } \right) \\
& \quad   + U\left[\left( {{\mathbf{p}}_i  \times {\mathbf{p}}_f } \right) \cdot {\mathbf{V}} - k^3 \left( {{\mathbf{V}} \times {\mathbf{p}}_f } \right) \cdot \hat z + V^3 \left( {{\mathbf{p}}_i  \times {\mathbf{p}}_f } \right) \cdot \hat z\right] \\
& \quad  \left. { + U^2 \left[ {\tfrac{1}
{2}\left( {k^3 } \right)^2  - \left( {p_i  \cdot p_f  + 2{\mathbf{p}}_i  \cdot {\mathbf{p}}_f  - m^2 } \right)} \right]} \right\} \\
\end{split}
\end{equation}

\subsection{Non-relativistic limit}
To make the above expression more informative, we consider its
behaviour in the non-relativistic limit, which gives us a clearer
physical insight and an ability to compare with the results of
other works.

If the incident electron is moving non-relativistically ($v\ll
1$), then the emitted photon will be soft ($\omega \rightarrow
0$). The non-relativistic limit of (\ref{20}) is most easily
obtained if we go back to (\ref{17}), and use it to rewrite the
unpolarized cross-section in terms of $V^\mu$ and $U$. Then, we
need the following approximate forms when $v \ll 1$:
\begin{equation}
\label{201}
\begin{aligned}
    & {p \cdot k} \cong \omega m,&
    & {\mathbf{R}} \cdot {\mathbf{p}}_f
    \cong {\mathbf{R}} \cdot
    {\mathbf{p}}_i \cong p_i^ \bot p_f^ \bot \sin \varphi _{if}\\
    & {V^\mu} \cong \frac{\left|
    \mathbf{q} \right|^2}{m \omega}\epsilon^{\mu *},&
    &{U} \cong 0,\\
    & \left|s\right|\cong\left|r\right|\cong 1,&
    & \frac{1+rs}{2rs}\cong\left\{ {\begin{array}{*{20}c}
   1, & \mathrm{if~} r=s  \\
   0, & \mathrm{if~} r\neq s  \\
 \end{array} } \right\}\equiv \Theta(rs),\\
    & \frac{E_i E_f-m^2}{{(p_{i}^{\bot}})^2}\cong\frac{1}{2}\left(1+\xi^2\right)
\end{aligned}
\end{equation}
where $\xi=\sqrt{1-\omega/(E_i-m)}$. Upon substituting these
relations in (\ref{20}), and keeping only the dominant terms, we
get for the various states of polarizations:
\begin{subequations}
\label{202}
\begin{align}
\mathop {\lim }\limits_{v \to 0} \frac{{d\sigma _{\left( {s \to
r,\sigma } \right)} }} {{d\omega d\Omega _\gamma  d\varphi _f }}&
= \frac{{e^2 \left( {\pi \delta } \right)^2 v}} {{32\pi ^4 m\omega
}}\Theta \left( {rs} \right)\left[ {\left( {1 + \xi ^2 } \right) +
2\xi \cos \left( {\varphi _{fk}  + \varphi _{ik} } \right)}\right]\\
\mathop {\lim }\limits_{v \to 0} \frac{{d\sigma _{\left( {s \to
r,\pi } \right)} }} {{d\omega d\Omega _\gamma  d\varphi _f }}& =
\frac{{e^2 \left( {\pi \delta } \right)^2 v}} {{32\pi ^4 m\omega
}}\Theta \left( {rs} \right)\cos ^2 \theta _k \left[ {\left( {1 +
\xi ^2 } \right) - 2\xi \cos \left( {\varphi _{fk}  + \varphi
_{ik} } \right)} \right]\\
\mathop {\lim }\limits_{v \to 0} \frac{{d\sigma _{\left( {s \to
r,\ell } \right)} }} {{d\omega d\Omega _\gamma  d\varphi _f }}& =
\frac{{e^2 \left( {\pi \delta } \right)^2 v}}
{{64\pi ^4 m\omega }}\Theta \left( {rs} \right)\left[ {\left( {1 + \xi ^2 } \right)\left( {1 + \cos ^2 \theta _k } \right)} \right. \hfill \nonumber\\
   & \left. { + 2\xi \cos \left( {\varphi _{fk}  + \varphi _{ik} } \right)\sin ^2 \theta _k  - 2\ell \left( {\frac{{2\omega }}
{{mv^2 }}} \right)\cos \theta _k } \right]
\end{align}
\end{subequations}
Some observations on these results are in order here:
\begin{description}
\item[1.] The factor $1/\omega$ implies the well-known infrared
catastrophe in the case of a soft photon $\omega\rightarrow 0$.
\item[2.] The appearance of the step function $\Theta(rs)$ means
that the spin projection of the electron is conserved in the
process of emission. But notice from (\ref{20}), that at higher
energies spin-flip is not prohibited. \item[3.] For
$\pi$-polarized states, we see the factor $\cos^2\theta_k$, which
is typical in this limit.
\end{description}

Expressions analogous to the first two equations in (\ref{202})
have been arrived at by Audretsch \textit{et al.} \cite{aud96},
using Dirac particles, and to the third by Gal'tsov and Voropaev
\cite{gal90}, using Klein-Gordon particles. Upon comparison, it is
seen that Audretsch's results readily reduce to (\ref{202}a) and
(\ref{202}b) by setting $\delta$ to zero everywhere except in the
$\sin^2(\pi\delta)$ factor, which appears in our case as
$\left(\pi\delta\right)^2$. Similar steps applied to Gal'tov's
result (which is quoted as a differential cross-section integrated
over azimuthal angles), reduces to our result for circular
polarization when we integrate it over azimuthal angles. In
comparing with Gal'tsov, however, one should note that the term
$\sim \xi \cos\left(\varphi_{fk}+\varphi_{ik}\right)$ is not
present their result, as this term appears as a result of the
inclusion of the spin degree of freedom, a fact that was noted
also in \cite{aud96}. Expectedly, our approach causes some
information loss in the non-relativistic limit, when compared to
Audretsch's and Gal'tsov's results, but it does agree in the
general form and characteristics. The differences lie in details
of how spin, flux, and momentum enter the cross-section. For
instance, Audretsch's result predicts a certain asymmetry in the
effect of the spin state due to the interaction between spin and
the magnetic field. Obviously this is lacking in our cross-section
at this limit. However, when the ``exact'' results are expanded in
terms of $\delta$ with the lowest non-vanishing order kept, they
coincide exactly with our expressions.

\section{Pair-production in the AB potential}
\subsection{Amplitude and Cross-section}
Turning now to the process of pair-production, we start with the
matrix element:
\begin{equation}
\label{101}
 S_{fi} = \left\langle {e^ - e^ + } \right|\hat S\left|
\gamma \right\rangle
\end{equation}
The lowest non-vanishing order of this amplitude is given by the
sum of the two diagrams of figure (\ref{fig2}):
\begin{figure}[htbp]
\includegraphics[width=8cm]{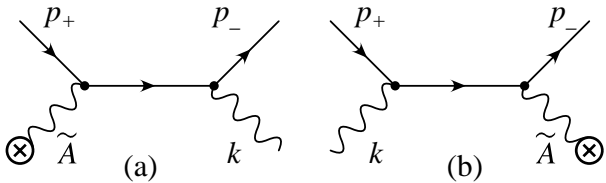}
\caption{\label{fig2} Tree-level diagrams for pair-production.}
\end{figure}

The processes of pair-production and bremsstrahlung are connected
by a crossing symmetry, which enables us to get the expressions
describing pair-production by appropriately modifying those for
bremsstrahlung. The matrix element and the subsequent calculations
for pair-production shall be conducted by transforming those of
bremsstrahlung according to the following rules:
\begin{equation}
\label{102}
\begin{array}{*{20}c}
 {p_i \leftrightarrow -p_+ ,} & {p_f \leftrightarrow p_-,}
 & {k\leftrightarrow -k,} & {\epsilon ^* \leftrightarrow \epsilon. } \\
 \end{array}
\end{equation}
where $p_{-}$, $p_{+}$ and $k$ are the momenta of the electron,
positron, and pair-producing photon, respectively. Accordingly, we
have: $q^\mu = p_{-}^{\mu}+p_{+}^{\mu}-k^\mu$ (other quantities
used here without definition retain the definitions used in
connection with bremsstrahlung). The sign difference between the
matrix element obtained by the crossing symmetry, and that
obtained using the process's own diagrams is due to the phase
convention, and can be safely dropped \cite{Pesk95}. We may thus
write down the matrix element for pair-production in the AB
potential immediately using (\ref{8}):
\begin{eqnarray}
\label{103}
 S_{fi}^{(2)}=\Lambda\bar{u}^{(\alpha)}(p_-)\Big[\not\!
\epsilon^{(\lambda)}\frac{\not\! p_- - \not\! k +m}{p_-\cdot k}\not\!\! R
+\not\!\! R\frac{\not\! k - \not\! p_+ +m}{p_+\cdot k}\not\!
\epsilon^{(\lambda)}\Big]v^{(\beta)}(p_+)
\end{eqnarray}
where $u^{(\alpha)}$ and $v^{(\beta)}$ are free-particle
``positive-energy'' and ``negative-energy'' solutions of Dirac's
equation, respectively, with indices $\alpha$ and $\beta$
referring to polarization states. The corresponding differential
cross-section is obtained by the formula:
\begin{eqnarray}
\label{104}
d\sigma_{(\lambda\rightarrow\alpha,\beta)}=\frac{|\Lambda|^2}{2\omega}
\Big(\frac{md^3p_-}{p_{-}^{0}(2\pi)^3}\Big)\Big(\frac{md^3p_+}{p_{+}^{0}(2\pi)^3}
\Big)
\operatorname{tr} \left\{ {\mathcal{P} _+^{(\beta)} \bar
F^{(\lambda)}\mathcal{P} _-^{(\alpha)} F^{(\lambda)}} \right\}
\end{eqnarray}
As in the bremsstrahlung case, we start with evaluating this
expression by neglecting the polarization of the created
particles, in which case we sum over polarization states $\alpha$
and $\beta$. The projectors in this case, which project over
states of positive- and negative-energy, respectively, are:
\begin{equation}
\label{105} \bar{\mathcal{P}}_-=\frac{\not\!
p_-+m}{2m},~~\bar{\mathcal{P}}_+=\frac{-\not\! p_++m}{2m}
\end{equation}
Plugging these into (\ref{104}), and doing steps analogous to what
was done to get (\ref{11}), we end up with:
\begin{subequations}
\label{106}
\begin{equation} {\frac{{d\bar{\sigma} _{(\lambda )} }} {{dp_{+}^{0} dp_{+}^{3} d\varphi
_{+} d\varphi _{-} }}}=\tilde{A}+\tilde{B} \label{subeq:106a}
\end{equation}
where:
\begin{equation}
\begin{aligned}
 \tilde{A}& = \frac{{-4e^2 (\pi \delta )^2 }}
{{(2\pi )^4 \left| {\mathbf{q}} \right|^4 \omega}}\left|
{\frac{{(\epsilon ^{(\lambda )} \cdot p_- )(R \cdot p_+ )}}
{{p_- \cdot k}}} \right. 
 \left. { + \frac{{(\epsilon ^{(\lambda )} \cdot p_+ )(R \cdot p_- )}}
{{p_+ \cdot k}} - R \cdot \epsilon ^{(\lambda )} } \right|^2 \hfill \\
  \tilde{B}& = \frac{{-e^2 (\pi \delta )^2 }}
{{2(2\pi )^4 \omega}}\left[ {\left| {\frac{{\epsilon ^{(\lambda )}
\cdot p_+ }} {{p_+  \cdot k}} - \frac{{\epsilon ^{(\lambda )}
\cdot p_- }}
{{p_-  \cdot k}}} \right|^2 } \right. 
  \left. { - \frac{{\left( {k^ \bot  } \right)^2 }}
{{(p_+  \cdot k)(p_-  \cdot k)}}} \right] 
\end{aligned}
\label{subeq:106b}
\end{equation}
\end{subequations}

We will now look at the differential cross-section when the
polarization of the created particles is considered, using the
$\hat{S}^3$ operator defined in (\ref{12}). The projector for the
electron was found in (\ref{15}). The positron's wavefunction
$\psi_{j}^{c}$ is obtained by charge-conjugating the wavefunction
in (\ref{14}), from which the spinor part is to be used to
construct the projector $\psi_{j}^{c}=i\gamma^{2}\psi_{j}^{*}$.
The desired projector in this case is constructed by finding the
product $u_{j}^{c}\bar{u}_{j}^{c}$, where , thus:
\begin{equation}
\label{107} {\mathcal{P}}_{+}^{(s)}(p)= \frac{-1} {{4sm}}\left( {s
- \gamma ^5 \gamma ^3 - \frac{{p^3 }} {m}\gamma ^5 } \right)\left(
{-\not\! p + m} \right)
\end{equation}

The polarized cross-section is obtained by calculating the trace
in (\ref{104}) using the projectors given in (\ref{107}) and
(\ref{15}). No actual calculation needs to be done to evaluate
this trace; we rather transform the results from bremsstrahlung as
given in equations (\ref{16}) and (\ref{17}). We skip the full
expression, which is essentially a repetition of the
bremsstrahlung result, and head straight for the cross-section for
a linearly polarized normally incident photon. We simply replace
each quantity $X$ in (\ref{16}) and (\ref{17}) by its transformed
counterpart $\tilde{X}$, in particular:
\begin{align}
\label{108}
 \tilde{a}& \equiv\frac{R\cdot p_+}{p_{-}\cdot
k}-\frac{R\cdot p_-}{p_{+}\cdot k},& \tilde{b}& \equiv
\frac{|\mathbf{q}|^2}{2}\frac{1}{p_-\cdot k},& \tilde{c}& \equiv
\frac{|\mathbf{q}|^2}{2}\frac{1}{p_+\cdot k}.
\end{align}
Doing that, we end up with:
\begin{equation}
\label{109}
\begin{split}
  \frac{{d\sigma _{\left( {\lambda  \to s,r} \right)} }}
{{dp_ + ^0 dp_ + ^3 d\varphi _ +  d\varphi _ -  }}& = \frac{{r +
s}} {{4s}}\left( {\frac{{d\bar \sigma _{\left( \lambda  \right)}
}}
{{dp_ + ^0 dp_ + ^3 d\varphi _ +  d\varphi _ -  }}} \right) \hfill \\
   & - \frac{{e^2 (\pi \delta )^2 }}
{{(2\pi )^4 sr\omega \left| {\mathbf{q}} \right|^4 }}\left\{ {\left( {\tilde V^3 } \right)^2 \left( {m^2  + p_ +   \cdot p_ -  } \right) - \left( {2 - s^2 } \right)\tilde U\tilde V^3 \left( {{\mathbf{p}}_ +   \times {\mathbf{p}}_ -  } \right) \cdot \hat z} \right. \hfill \\
   & - \tilde Us^2 \left[ {\left( {{\mathbf{p}}_ +   \times {\mathbf{p}}_ -  } \right) \cdot {\mathbf{\tilde V}} + p_ + ^3 \left( {{\mathbf{\tilde V}} \times {\mathbf{p}}_ -  } \right) \cdot \hat z + p_ + ^3 \left( {{\mathbf{\tilde V}} \times {\mathbf{p}}_ +  } \right) \cdot \hat z} \right] \hfill \\
  & \left. { + \tilde U^2 \left[ {2\left( {p_ + ^3 } \right)^2  + p_ +   \cdot p_ -   + 2{\mathbf{p}}_ +   \cdot {\mathbf{p}}_ -   + m^2 } \right]} \right\} \hfill \\
\end{split}
\end{equation}
where $s$ and $r$ are the eigenvalues of $\hat{S}^3$ for the
electron and positron, respectively, and $\tilde V^\mu   \equiv
\tilde a\epsilon ^\mu   + \eta ^\mu ( {\tilde b - \tilde c})$, and
$\tilde U \equiv  - \epsilon ^\mu  ( {\tilde b + \tilde c})$.

We notice from (\ref{109}) and the fact that the photon is assumed
normally incident on the solenoid, that the cross-section for
pair-production from an unpolarized photon is more similar to that
from a $\pi$-polarized photon than to a $\sigma$-polarized photon.

\subsection{Low photon-energy limit}
Under the low photon energy limit the energy carried by the
pair-producing photon is just above the pair-production threshold:
\begin{equation}
\label{110} \omega  \gtrsim 2m
\end{equation}
As a result, the energy of each one of the created particles is
almost its rest energy, so that the motion is non-relativistic,
and the following approximations hold:
\begin{align}
\label{}
    & \omega \gg \left|\mathbf{p}\right|,
    & p \cdot k \cong 2m^2,\nonumber\\
    & R \cdot X \cong - k^2 X^1 + k^1 X^2,
    & {\tilde{V}}^\mu \cong 0,\nonumber\\
    & \tilde{U}\cong
    -\frac{\left|\mathbf{q}\right|^2}{2m^2}\epsilon^3
\end{align}
Substituting these into the cross-section formula, we get after
tidying up:
\begin{equation}
\label{111} \mathop {\lim }\limits_{\omega  \to 2m} \frac{{d\sigma
_{\left( {\lambda  \to s,r} \right)} }} {{dp_ + ^0 dp_ + ^3
d\varphi _ + d\varphi _ -  }} = \frac{{e^2 (\pi \delta )^2 }}
{{128\pi ^4 m^3 }}\left( {1 - rs} \right)\left| {\epsilon ^3 }
\right|^2
\end{equation}
Here also, it is possible to make several observations:
\begin{description}
\item[1.]The angular and spectral distributions of the
non-relativistic created particles are uniform with respect to all
the parameters in which the cross-section is differential.
\item[2.]In this limit, due to the factor $(1-rs)$, the created
pair have opposite signs of spin projection. \item[3.] When
$k^3=0$, we have $\epsilon^3=0$ for a $\sigma$-polarized photon
and $\epsilon^3=1$ for a $\pi$-polarized photon, so that the
particles should in this limit be predominantly created by
$\pi$-polarized photons. \item[4.] Had we used circular
polarization, we see that in this limit, the two states of
polarization are indistinguishable.
\end{description}

An expression similar to that in equation (\ref{111}) was arrived
at by Skarzhinsky \emph{et al.} \cite{skar96}. As was in the
bremsstrahlung case, the two expressions agree in form and general
features, but differ in the details they convey. In particular,
the expression of Skarzhinsky \emph{el al.} has a structure with
more complicated dependence on spin, flux and momentum. As with
the bremsstrahlung case, these effects are turned off by setting
$\delta$ to zero everywhere except in the $\sin^2(\pi\delta)$
factor, effectively keeping the leading term in an expansion in
powers of $\delta$, in which case Skarzhinsky's expression reduces
to ours. Aside from this point, the above observations confirm
those in \cite{skar96}, except for the last observation, which was
not mentioned there.

\section{Conclusion}
In this paper we have studied the QED processes of bremsstrahlung
and pair-production in the AB potential for Dirac particles. We
have used the formalism of covariant perturbation theory to the
lowest non-vanishing order in the coupling constant. The matrix
elements were found, and the corresponding differential
cross-section formulae were calculated, and the effect of
polarization for both Dirac particles and the photon was taken
into account.

We have confirmed in our work the main results that were formerly
arrived at using the exact wavefunction method as used in
\cite{aud96,skar96}, and also observed the differences and
similarities with the spinless case as investigated in
\cite{gal90}. In particular, we have compared the expressions for
the differential cross-sections in some limiting cases, and have
seen that the results coincide when the exact result is expanded
to the same order in the fraction of flux $\delta$, but with an
expected loss in details, pertaining especially to spin.

There are two interesting problems related to the work presented
in this paper, both of which are currently under investigation.
The first problem is to conduct a partial wave analysis of the two
processes at hand. The reason why this is especially interesting
is that, unlike the calculations in \cite{aud96,skar96}, which are
to first order in $\alpha$, our work is to second order, and that
means that we have to deal with a propagator in our calculation.
Moreover, aside from the fact that this calculation is of
intrinsic interest, the former works, upon conducting partial wave
analysis, have reported a selection rule that prohibits the
incoming and outgoing particles to have angular momentum
projections in the same direction.

The second problem is to conduct the same calculations done here
but using spinless Klein-Gordon particles. We expect that this may
involve a well-known difficulty \cite{fein63,cor78}, whereby a
discrepancy between the result of exact wavefunction method and
that obtained with the first Born approximation could manifest
itself.

\begin{acknowledgments}
The authors wish to thank Mr. F. Shahin for his valuable help in
producing the Feynman diagrams appearing in this work.
\end{acknowledgments}

\bibliography{paper}

\end{document}